\documentclass[showpacs,article]{revtex4}
\usepackage{epsfig}
\def\beq{\begin{equation}}
\def\eeq{\end{equation}}
\def\beqa{\begin{eqnarray}}
\def\eeqa{\end{eqnarray}}

\def\GeV{\nobreak\,\mbox{GeV}}

\def\pli{p^\prime}

\begin{document}
\title{$D^* D_s K$ and $D_s ^* D K$ vertices in a QCD Sum Rule approach}
\author{M.E. Bracco,  A. Cerqueira Jr., M. Chiapparini, A. Loz\'ea\footnote{Permanent address: {\em Intituto de F\'{\i}sica, Universidade Federal do Rio de Janeiro, C.P. 68528, 21941-972 Rio de Janeiro, RJ, Brazil.}}, }
\affiliation{Instituto de F\'{\i}sica, Universidade do Estado do Rio de 
Janeiro, 
Rua S\~ao Francisco Xavier 524, 20550-900 Rio de Janeiro, RJ, Brazil}
\author{M. Nielsen}
\affiliation{Instituto de F\'{\i}sica, Universidade de S\~{a}o Paulo, 
C.P. 66318, 05389-970 S\~{a}o Paulo, SP, Brazil.}

\begin{abstract}
We calculate the strong form factors and coupling constants of 
$ D^* D_s K$ and $D_s^* D K$ vertices using the QCD sum rules technique. In each case 
we have  considered two different cases for the off-shell particle
in the vertex: the ligthest meson and one of the heavy mesons.
The method gives the same coupling constant for each vertex. When the 
results for different vertices are compared, they show that the SU(4)
symmetry is broken by around 40\%.

\end{abstract}

\pacs{14.40.Lb,14.40.Nd,12.38.Lg,11.55.Hx}

\maketitle

The knowledge of the form factors in hadronic vertices is of crucial 
importance to estimate hadronic amplitudes when hadronic degrees of freedon
are used. When all the particles in a hadronic vertex are on-mass-shell,
the effective fields of the hadrons describe pointlike physics. However,
when at least one of the particles in the vertex is off-shell, the finite 
size effects of the hadrons become important.

In this work we study the $D^* D_s K$ and $D_s ^* D K$ vertices, which are
fundamental to the evaluation of the dissociation cross section of 
$J/\Psi$ by kaons  when using effective Lagrangians.
The supression of charmonium production is one of the most tradicional
signatures of the quark-gluon plasma (QGP) formation in relativistic heavy
ion collisions \cite{matsui}. The dissociation of charmoniums in the QGP 
due to color screening would lead to a reduction of its production in 
such collisions. However, using the charmonium suppression as a 
signature of QGP formation requires the understanding of $J/\Psi$ 
production and absortion mechanisms in hadronic matter, because this 
suppression may be due to the interactions with the comovers during the 
collision \cite{cassing,armesto}.

One of the approaches used to study the interaction of charmonium with 
the hadronic medium, mainly in the low energy region ($\sqrt{s}<10\GeV$), 
is based on effective SU(4) Lagrangians 
\cite{mamu,lin,haglin,oh,nnr,regina}. This technique, however, requires 
the detailed knowledge of the  form factors in the hadronic
vertices. The calculated cross section may change by a factor of two if a 
soft, instead of a hard, form factor is used in the vertices containing
charmed mesons. 

This situation gave us the motivation to start a program to calculate
charmed form factors and coupling constants, using the QCD sum rules 
approach \cite{svz}. We have been continuously working on this problem 
and computing different vertices 
\cite{ddpi,ddpir,ddrho,matheus1,matheus2,dasilva,psid*d*}. An interesting 
subproduct of such calculations, \cite{ddpir,ddrho,psid*d*,matheus1}, was the 
understanding of the behavior of the off-shell particle probing of the 
vertex: heavier particles resolves better the structure of the vertex, 
while lighter particles are more suitable for measuring its size. 
This conclusion is also supported in the present work. 

As a part of this project we evaluate, in the present calculation,
the form factors in the vertices $D^* D_s K$ and $D_s ^* D K$, and 
compare the results with the predictions from the exact SU(4) symmetry 
\cite{regina}.

Following the QCDSR formalism described in our previous works 
\cite{ddpi,ddpir,ddrho,matheus1,matheus2,dasilva,psid*d*}, we write the three-point 
correlation function associated 
with the $D^* D_s K$  vertex, which is given by
\begin{equation}
\Gamma_{\mu}^{(K)}(p,\pli)=\int d^4x \, d^4y \;\;
e^{i\pli\cdot x} \, e^{-i(\pli-p)\cdot y}
\langle 0|T\{j_{\mu}^{D^*}(x) {j^K}^\dagger(y) 
 {j^{D_s}}^\dagger(0)\}|0\rangle  \label{corkoffnos} 
\end{equation}
for $K $ meson off-shell, where the interpolating currents are 
$j_\mu^{D^*}=\bar c\gamma_\mu d$,
$j^K=i\bar s\gamma_5 d$ and
$j^{D_s}=i\bar c\gamma_5 s$, and
\begin{equation}
\Gamma_{\mu \nu}^{(D_s)}(p,\pli)=\int d^4x \, 
d^4y \;\; e^{i\pli\cdot x} \, e^{-i(\pli-p)\cdot y}\;
\langle 0|T\{j_{\mu}^{K}(x)  {j^{D_s}}^\dagger(y) 
 {j_{\nu}^{D^*}}^\dagger(0)\}|0\rangle \label{cordsoffnos} 
\end{equation}
for $D_s$ meson off-shell, with the interpolating currents 
$j^{K}_\mu = \bar u \gamma_{\mu}\gamma_{5} s$, 
$j^{D_s} = i \bar c \gamma_{5} s$, 
$j_\mu^{D^*}= \bar u \gamma_{\mu} c$, with
$u$, $d$, $s$ and $c$ being the $up$, $down$, $strange$ and $charm$ 
quark fields respectively. In both cases, each one of these currents has 
the same quantum numbers as the corresponding mesons.

Using the above currents to evaluate the correlation functions 
(\ref{corkoffnos}) and (\ref{cordsoffnos}), the theoretical or QCD side 
is obtained.  The framework to calculate the correlators in the QCD side 
is the Wilson operator product expansion (OPE). The Cutkosky's rule 
allows us to obtain the double discontinuity of the correlation function 
for each one of the Dirac structures appearing in the correlation function.
Calling $\rho_i$ the spectral density for the Dirac structure $i$,
we can write the correlation function  as a double dispersion relation 
over the virtualities $p^2$ and ${\pli}^2$, holding $Q^2= -q^2$ fixed. 
Therefore, the amplitudes $\Gamma_i$ are given by: 
\begin{equation}
\Gamma_i(p^2,{\pli}^2,Q^2)=-\frac{1}{\pi^2}\int_{s_{min}}^{s_0} ds
\int_{u_{min}}^{u_0} du \:\frac{\rho_i(s,u,Q^2)}{(s-p^2)(u-{\pli}^2)},
\label{dis}
\end{equation}
where the spectral density $\rho_i(s,u,Q^2)$ equals the double 
discontinuity of the amplitude
$\Gamma_i(p^2,{\pli}^2,Q^2)$. The amplitudes receive contributions  from 
all terms in the OPE. The leading contribution comes from 
the perturbative term, shown in Fig.~\ref{fig1}.
\begin{figure}[t]
\begin{picture}(12,3.5)
\put(0.0,0.5){\vector(1,0){1.5}}
\put(3.5,0.5){\vector(-1,0){2}}
\put(3.5,0.5){\vector(1,0){1.5}}
\put(1.5,0.5){\vector(1,1){1}}
\put(2.5,1.5){\vector(1,-1){1}}
\put(2.5,3){\vector(0,-1){1.5}}
\put(2.65,2.75){$q_\alpha$}
\put(0.25,0.65){$p_\mu$}
\put(4.55,0.65){$p'_\nu$}
\put(2.4,0.2){$\bar c$}
\put(1.85,1.1){$s$}
\put(3,1.1){$d$}
\put(2.4,1.2){$y$}
\put(1.75,0.53){$0$}
\put(3.05,0.53){$x$}
\put(1.95,2.2){$K^0$}
\put(0.35,0.1){$D_s^-$}
\put(4,0.1){$D^{*-}$}
\put(7,0.5){\vector(1,0){1.5}}
\put(10.5,0.5){\vector(-1,0){2}}
\put(10.5,0.5){\vector(1,0){1.5}}
\put(8.5,0.5){\vector(1,1){1}}
\put(9.5,1.5){\vector(1,-1){1}}
\put(9.5,3){\vector(0,-1){1.5}}
\put(9.65,2.75){$q_\alpha$}
\put(7.25,0.65){$p_\mu$}
\put(11.55,0.65){$p'_\nu$}
\put(9.4,0.2){$\bar u$}
\put(8.85,1.1){$c$}
\put(10,1.1){$s$}
\put(9.4,1.2){$y$}
\put(8.75,0.53){$0$}
\put(10.05,0.53){$x$}
\put(8.75,2.2){$D_s^-$}
\put(7.35,0.1){$D^{*0}$}
\put(11,0.1){$K^-$}
\end{picture}
\caption{Perturbative diagrams for $K$ off-shell (left) and $D_s$
off-shell (right) correponding to the $D^*D_sK$ vertex.}
\label{fig1}
\end{figure}
The phenomenological side of the sum rule, which is written in terms of 
the mesonic degrees of freedom, is parametrized in terms  of the form 
factors, meson decay constants and meson masses. The QCD sum rule is 
obtained by matching both representations, using the universality 
principle. The matching is improved by performing a double Borel 
transform on both sides.

The perturbative contribution for both Eqs.~(\ref{corkoffnos}) and 
(\ref{cordsoffnos}), written in terms of Eq.(\ref{dis}), is given by
\beqa
\rho^{(K)}_{\mu}(s,u,Q^2)& = & \frac{3}{2\pi\sqrt\lambda}
 \left\{ p_{\mu} \left[A \left(m_c^2-m_c m_s - 2k\cdot p+p\cdot\pli\right)
 +2\pi\left(m_c^2-k\cdot\pli\right) \right]\right.  \nonumber \\
&&+ \left. \pli_{\mu}\left[B\left(m_c^2-m_c m_s - 2k\cdot p+p\cdot\pli
\right)
+2\pi\left(-m_c^2+m_cm_s+k\cdot p\right) \right] \right\}
\label{sdkoff}
\eeqa
for $K$ off-shell, and 
\beqa
\rho^{(D_s)}_{\mu \nu}(s,u,Q^2)&=&-\frac{3 i}{2\pi\sqrt\lambda} 
\left\{ 
g_{\mu \nu} \left[\pi\left(m_s\left(s-m_c^2\right)-m_c\left(u-m_s^2\right)
\right) 
+ 2 D\left(m_s-m_c\right)\right]\right.  \nonumber \\
&&+ \left(p_{\mu}\pli_{\nu}+\pli_{\mu}p_{\nu}\right)\left[A m_c - B m_s+2 
C(m_s-m_c)  \right]\nonumber \\
&&+\left. p_{\mu} p_{\nu}2 \left[F(m_s-m_c)- A m_s \right]
+\pli_{\mu} \pli_{\nu} 2\left[B m_c+E(m_s-m_c)\right] \right\}
\label{sddsoff}
\eeqa
for $D_s$ off-shell. Here $s=p^2$, $u=p'^2$, $t=-Q^2$, 
$\lambda\equiv\lambda(s,t,u) = 
s^2+t^2+u^2-2st-2su-2tu$, $k\cdot p= \frac{s+m_c^2-m_s^2}{2}$, 
$p\cdot\pli=\frac{s+u-t}{2}$, $k\cdot\pli= \frac{u+m_c^2}{2}$, and $A$, 
$B$, $C$, $D$, $E$ and $F$ are functions of $\{s,t,u\}$, given by the 
following expressions:
\[
\begin{array}{rclrcl}
A&=&\displaystyle{\frac{2\pi}{\sqrt{s}}\left(\overline{k}_0-\frac{|
\overline{\vec{k}}|p'_0}{|\vec p'|}
\cos \overline{\theta} \right)},&
B&=&\displaystyle{2 \pi \frac{|\overline{\vec{k}}|}{|\vec p'|}\cos\
\overline{\theta}}, \\
C&=&\displaystyle{\frac{\pi{\overline{k}_0}^2}{\sqrt{s}|\vec{\pli}|}
\left[2 \cos\overline{\theta} - \frac{p'_0}{|\vec p'|}\left(3\cos^2
\overline{\theta} -1 \right) \right]},\;\;\;\;&
D&=&\displaystyle{\pi\overline{k}_0^2 \left(1-\cos^2\overline{\theta}
\right)}, \\
E&=&\displaystyle{\frac{\pi\overline{k}_0^2}{|\vec p'|^2}\left(3\cos^2
\overline{\theta}-1\right)},&
F&=&\displaystyle{\frac{\pi\overline{k}_0^2 }{s}\left[3 -\frac{4p'_0}{|
\vec p'|}\cos\overline{\theta}
+\frac{{p'_0}^2}{|\vec p'|^2}\left(3\cos^2 \overline{\theta}-1\right)-
\cos^2\overline{\theta}\right]},\\
p'_0&=&\displaystyle{\frac{s+u-t}{2\sqrt{s}}},&|\vec p'|^2&=&
\displaystyle{\frac{\lambda}{4s}},
\end{array}\label{abcdef}
\]
where
\[
\begin{array}{rclrclrcl}
\overline{k}_0&=&\displaystyle{\frac{s+m_c^2-m_s^2}{2\sqrt{s}}},\;\;\;\;&
|\overline{\vec{k}}|&=&\displaystyle{\sqrt{\overline{k}_0^2-m_c^2}},\;\;
\;\;&
\cos\overline{\theta}&=&\displaystyle{\frac{2p'_0\overline{k}_0-m_c^2-u}
{2|\vec{\pli}||\overline{\vec{k}}|}},
\end{array}
\]
for $K$ off-shell, and
\[
\begin{array}{rclrclrcl}
\overline{k}_0&=&\displaystyle{\frac{s-m_c^2}{2\sqrt{s}}},\;\;\;\;&
|\overline{\vec{k}}|&=&\displaystyle{\overline{k}_0},\;\;\;\;&
\cos\overline{\theta}&=&\displaystyle{\frac{2p'_0\overline{k}_0+m_s^2-u}
{2|\vec{\pli}||\overline{\vec{k}}|}},
\end{array}
\]
for $D_s$ off-shell.

The phenomenological side of the vertex functions is obtained considering 
the contributions of the $D_s$ and $D^*$ mesons to the matrix element in 
Eq.~(\ref{corkoffnos}) and the $D^*$ and $K$ mesons to the matrix element 
in Eq.~(\ref{cordsoffnos}). We introduce the meson decay constants 
$f_{K}$, 
$f_{D_s}$ and $f_{D^*}$, which are defined by the following matrix 
elements:
\beqa
\langle 0|j^{K}|{K}\rangle&=&\frac{m_{K}^2 f_{K}}{m_s+m_q}, \label{fk} \\
\langle 0|j^{D_s}|{D_s}\rangle&=&\frac{m_{D_s}^2}{m_c+m_s}f_{D_s}, 
\label{fds} \\ 
\langle 0|j_{\nu}^{D^*}|{D^*}\rangle&=&m_{D^*}f_{D^*} \epsilon^*_{\nu},
\label{fd*}
\eeqa
where $\epsilon_{\nu}$ is the polarization vector of the $D^*$ meson. 

In principle, we can work with any Dirac structure appearing in the 
amplitude in Eqs.~(\ref{corkoffnos}) and (\ref{cordsoffnos}). 
However, there are some points that one must follow: (i) the chosen 
structure must also appear in the 
phenomenological side and (ii) the chosen structure must have a stability 
that guarantees a good match between the two sides of the sum rule. 
The structures that obey these two points are $\pli_{\mu}$, in the case
$K$ off-shell, and $\pli_{\mu}\pli_{\nu}$ in the case $D_s$ off-shell. 
The corresponding  phenomenological amplitudes in these structures are 
\begin{equation}
\Gamma^{(K)ph}(p^2,{\pli}^2,Q^2)=g^{(K)}_{D^*D_sK}(Q^2)
\frac{f_{D^*}f_{D_s}f_{K} m_{D^*}m_{D_s}^2m_K^2}
{(m_c+m_s)m_s(p^2-m^2_{D_s})({\pli}^2-m^2_{D^*})(Q^2+m^2_K)}
\left(1+\frac{m^2_{D_s}+Q^2}{m^2_{D^*}}\right) 
\label{phenkoff}
\end{equation}
for the $K$ off-shell, and 
\begin{equation}
\Gamma^{(D_s)ph}(p^2,{\pli}^2,Q^2)= g^{(D_s)}_{D^*D_sK}(Q^2)
\frac{(-2)if_{D^*}f_{D_s}f_{K}m_{D^*}m^2_{D_s}}
{ (m_c+m_s)(p^2-m^2_{D^*})({\pli}^2 -m^2_{K})(Q^2+m^2_{D_s})}
\label{phendsoff}
\end{equation}
for $D_s$ off-shell.

In the case of $K$ off-shell the contribution of the quark condensate 
vanishes after the double Borel transform. In the case of the $D_s$ 
off-shell, the quark condensate does not contribute to the chosen 
structure.

To write the sum rules we equate each phenomenological amplitude in 
Eqs.~(\ref{phenkoff})--(\ref{phendsoff}), with the expression obtained 
by substituting the corresponding spectral density in
Eqs.~(\ref{sdkoff})--(\ref{sddsoff}) into Eq.~(\ref{dis}). 
The matching between both sides is improved by performing a double Borel 
transformation \cite{io2} in the variables 
$P^2=-p^2\rightarrow M^2$ and $P'^2=-{\pli}^2\rightarrow M'^2$. We get 
then the final form of the sum rule, 
which allow us to obtain the form factors $g^{(M)}_{D^* D_s K}(Q^2)$ 
appearing in Eqs.~(\ref{phenkoff})--(\ref{phendsoff}), where $M$ stands
for the off-shell meson. 

We use Borel masses satisfying the constraint 
$M^2/M'^2=m_{in}^2/m_{out}^2$, where $m_{in}$ and 
$m_{out}$ are the masses of the incoming and outcoming meson respectively.
 In the case of the $K$ meson off-shell, this constraint gives
$M^2/M'^2=m_{D_s}^2/m_{D^*}^2$. For the $D_s$ meson off-shell, the 
relation should be $M^2/M'^2=m_{D^*}^2/m_K^2$. However, the small value 
of the $K$ mass spoils the stability of the Borel transformation. Thus, 
as is common in the literature, we change the $K$ mass for the $\rho$ 
mass. The resulting relation is then $M^2/M'^2=m_{D^*}^2/m_\rho^2$.

The values of the parameters used in the calculation of the $D^* D_s K$ 
vertex are  despicted in Table~\ref{param1}.
\begin{table}[t]
\begin{tabular}{|c|c|c|c|c|c|c|c|c|c|}\hline
$m_q$&$m_s$&$m_c$&$m_K$&$m_{D_s}$&$m_{D^*}$&$f_K$\cite{fk}&$f_{D_s}$
\cite{fds}&$f_{D^*}$\cite{fd*}\\ 
\hline\hline
0.0&0.13&1.2&0.498&1.97&2.01&0.160&0.280&0.240\\ \hline
\end{tabular}
\caption{Parameters used in the calculation of the QCD sum rule for the 
$D^* D_s K$ vertex. All quantities are in $\GeV$.}\label{param1}
\end{table}
The continuum thresholds $s_0$ and $u_0$, appearing in Eq.~(\ref{dis}), 
are given by $s_0=(m_{in} + \Delta_s)^2$ and $u_0=(m_{out}+\Delta_u)^2$, 
where $m_{in}$ and $m_{out}$ are the masses of the incoming and outcoming 
mesons respectively. For the $K$ off-shell we have $m_{in}=m_{D_s}$ and 
$m_{out}=m_{D^*}$, and for the $D_s$ off-shell we have $m_{in}=m_{D^*}$ 
and $m_{out}=m_K$ (see Fig.~\ref{fig1}).

\begin{figure}[b] 
\begin{center}
\epsfig{file=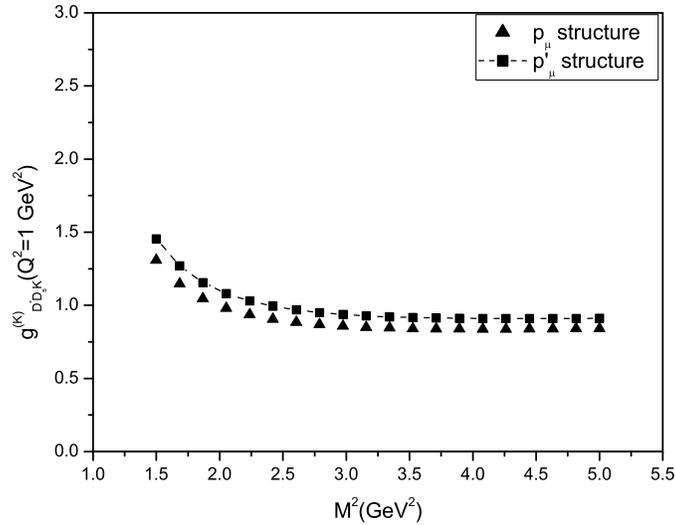}
\caption{Stability of $g^{(K)}_{D^*D_sK}(Q^2=1 \GeV^2)$, as a function of the 
Borel mass $M^2$.}
\label{fig2}
\end{center}
\end{figure}

\begin{figure}[t] 
\begin{center}
\epsfig{file=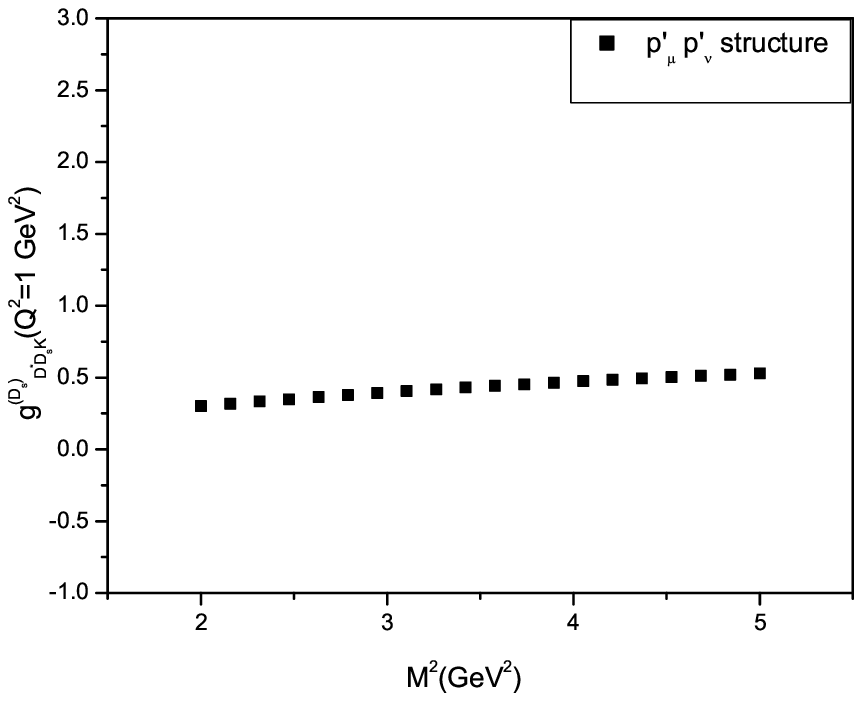}
\caption{Stability of $g^{(D_s)}_{D^*D_sK}(Q^2=1 \GeV^2)$, as a function of the 
Borel mass $M^2$.}
\label{fig3}
\end{center}
\end{figure}

\begin{figure}[t] 
\begin{center}
\epsfig{file=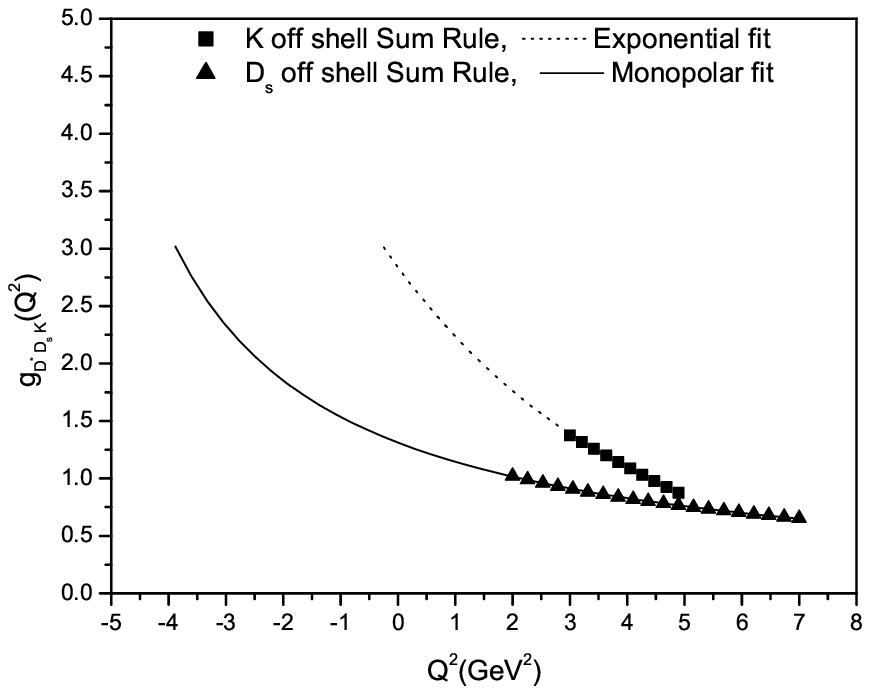}
\caption{$g^{(K)}_{D^*D_sK}$ (squares) and $g^{(D_s)}_{D^*D_s K}$ 
(triangles) form factors as a function of $Q^2$ from the QCDSR 
calculation of this work. The solid (dotted) line corresponds to the 
monopole (exponential) parametrization of the QCDSR results for each case.}
\label{fig4}
\end{center}
\end{figure}

Using $\Delta_s=\Delta_u = 0.5 \GeV $ for the continuum thresholds 
and fixing $Q^2=1\GeV^2$, we found a good stability of the form factor
$g_{D^*D_sK}^{(K)}$, as a function of the Borel mass $M^2$, in the 
interval $3< M^2 < 5 \GeV^2$, as can be seen in Fig.~\ref{fig2}. In the 
case of the form factor $g_{D^*D_sK}^{(D_s)}$ the interval for stability 
of the sum rule is $2<M^2<5 \GeV^2 $, as can be seen in Fig.~\ref{fig3}. 
Fixing $\Delta_s=\Delta_u=0.5 \GeV$ and $M^2=3 \GeV ^2$, we evaluate 
the momentum dependence of both form factors. The results are shown in 
Fig.~\ref{fig4}, where the squares corresponds to the 
$g_{D^*D_s K}^{(K)}(Q^2)$ form factor in the  interval where the 
sum rule is valid. The triangles are the result of the sum rule for the 
$g_{D^*D_s K}^{(D_s)}(Q^2)$ form factor. 
\begin{figure}[t] 
\begin{center}
\epsfig{file=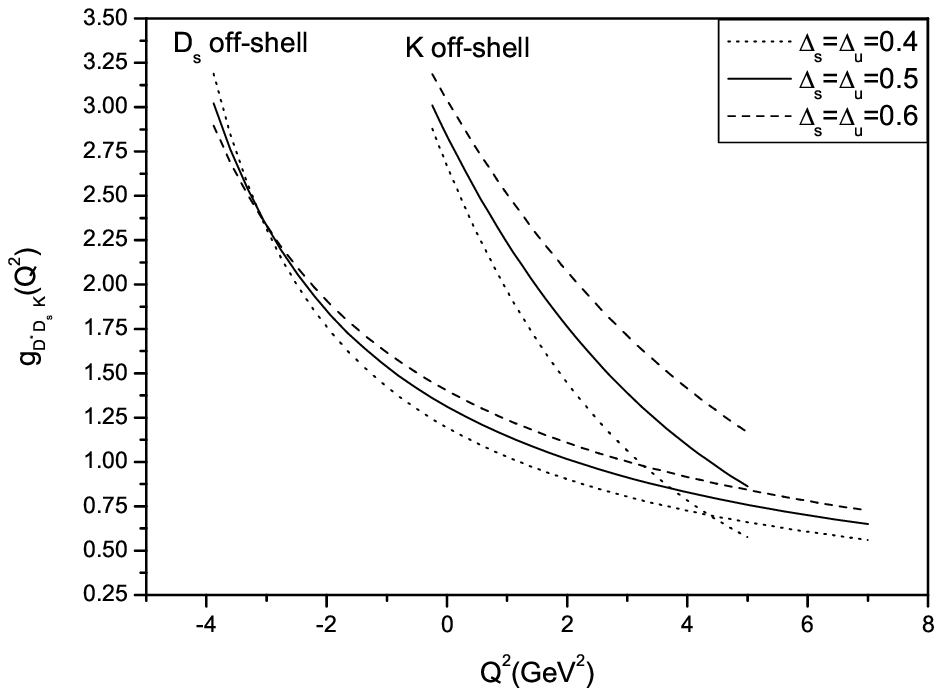}
\caption{Dependence of the form factor with the continuum threshold, for 
$K$ and $D_s$ off-shell cases. The dotted line correponds to $\Delta_s=
\Delta_u=0.4~\GeV$, the solid line corresponds to $\Delta_s=\Delta_u=0.5~
\GeV$ and  the dashed one corresponds to $\Delta_s=\Delta_u=0.6~\GeV$.}
\label{fig5}
\end{center}
\end{figure}
In the case of the $K$ meson off-shell, our numerical results can be 
parametrized by an exponencial function (dotted line in Fig.~\ref{fig4}):
\begin{equation}
g_{D^* D_s K}^{(K)}(Q^2)= 2.83 \; e^{-\frac{Q^2}{4.19}} .
\label{expnos}
\end{equation}
As in Ref.\cite{ddrho}, we define the coupling constant as the value of 
the form factor at  $Q^2=-m^2_M$, where $m_M$ is the mass of the 
off-shell meson. 
For the $K$ off-shell case the resulting coupling constant is:
\begin{equation}
g_{D^* D_s K}^{(K)}= 3.01.  \label{gexpnos}
\end{equation}
In the case when the $D_s$ meson is off-shell, our sum rule results can 
be parametrized by a monopole formula (solid line in Fig.~\ref{fig4}):  
\begin{equation}
g_{D^* D_s K}^{(D_s)}(Q^2)=\frac{9.01}{Q^2+6.86},
\label{mononos}
\end{equation}
giving the following coupling constant, obtained at the $D_s$ pole:
\begin{equation}
g_{D^* D_s K}^{(D_s)}= 3.02\label{gmononos}.
\end{equation}

Comparing the results in Eqs.(\ref{gexpnos}) and (\ref{gmononos}) we see 
that the method used to extrapolate the QCDSR results in both cases, 
$K$ and $ D_s$ off-shell, allows us to extract values for the coupling 
constant which are in very good agreement with each other.

In order to study the dependence of these results with the continuum 
threshold, we vary $\Delta_s=\Delta_u$ in the interval
$0.4 \le \Delta_s=\Delta_u \le 0.6~\GeV$, as can be seen in Fig~\ref{fig5}. 
This procedure give us uncertainties in such a way that the final results for the couplings in each case are: 
$g_{D^* D_s K}^{(K)}= 3.02 \pm 0.15 $ and $g_{D^* D_s  K}^{(D_s)}= 
3.03 \pm 0.14 $.

\begin{figure}[t]
\begin{picture}(12,3.5)
\put(0.0,0.5){\vector(1,0){1.5}}
\put(3.5,0.5){\vector(-1,0){2}}
\put(3.5,0.5){\vector(1,0){1.5}}
\put(1.5,0.5){\vector(1,1){1}}
\put(2.5,1.5){\vector(1,-1){1}}
\put(2.5,3){\vector(0,-1){1.5}}
\put(2.65,2.75){$q_\alpha$}
\put(0.25,0.65){$p_\mu$}
\put(4.55,0.65){$p'_\nu$}
\put(2.4,0.2){$\bar c$}
\put(1.85,1.1){$u$}
\put(3,1.1){$s$}
\put(2.4,1.2){$y$}
\put(1.75,0.53){$0$}
\put(3.05,0.53){$x$}
\put(1.95,2.2){$K^-$}
\put(0.35,0.1){${\overline{D}}^0$}
\put(4,0.1){$D_s^{*-}$}
\put(7,0.5){\vector(1,0){1.5}}
\put(10.5,0.5){\vector(-1,0){2}}
\put(10.5,0.5){\vector(1,0){1.5}}
\put(8.5,0.5){\vector(1,1){1}}
\put(9.5,1.5){\vector(1,-1){1}}
\put(9.5,3){\vector(0,-1){1.5}}
\put(9.65,2.75){$q_\alpha$}
\put(7.25,0.65){$p_\mu$}
\put(11.55,0.65){$p'_\nu$}
\put(9.4,0.2){$\bar s$}
\put(8.85,1.1){$c$}
\put(10,1.1){$u$}
\put(9.4,1.2){$y$}
\put(8.75,0.53){$0$}
\put(10.05,0.53){$x$}
\put(8.75,2.2){${\overline{D}}^0$}
\put(7.35,0.1){$D_s^{*+}$}
\put(11,0.1){$K^+$}
\end{picture}
\caption{Perturbative diagrams for $K$ off-shell (left) and $D$
off-shell (right) correponding to the $D_s^*DK$ vertex.}
\label{fig6}
\end{figure}
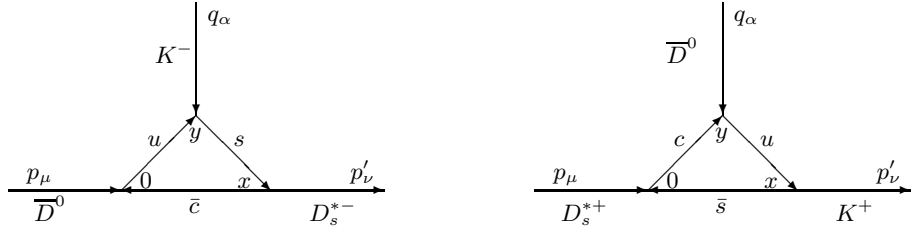

Now we study the $D_s^* D K$ vertex. The treatment is similar to 
the previus case. The correlation functions are 
\begin{equation}
\Gamma_{\mu}^{(K)}(p,\pli)=\int d^4x \, d^4y \;\;
e^{i\pli\cdot x} \, e^{-i(\pli-p)\cdot y}
\langle 0|T\{j_{\mu}^{D^*_s}(x) {j^{K}}^\dagger(y) 
 {j^{\bar D}}^\dagger(0)\}|0\rangle  \label{corkoffang} 
\end{equation}
for $K$ meson off-shell, where the interpolating currents are 
$j_\mu^{D_s^*}=\bar c\gamma_\mu s$,
$j^K=i\bar u\gamma_5 s$ and
$j^D=i\bar c \gamma_5 u$, and 
\begin{equation}
\Gamma_{\mu \nu}^{(D)}(p,\pli)=\int d^4x \, 
d^4y \;\; e^{i\pli\cdot x} \, e^{-i(\pli-p)\cdot y}\;
\langle 0|T\{j_\mu^K(x)  {j^D}^\dagger(y)
 {{j_\nu}^{D_s^*}}^\dagger(0)\}|0\rangle \label{cordoffang} 
\end{equation}
for $D$ meson off-shell, with the interpolating currents 
$j^{K}_\mu = \bar u \gamma_\mu\gamma_5 s$,
$j_{\nu}^{D^*_s}= \bar c \gamma_{\nu} s$, and 
$j^{D} = i \bar u \gamma_{5} c$. See Fig.~{\ref{fig6}} for understanding
the perturbative contribution with these currents.

For each correlation function, Eqs.~(\ref{corkoffang}) and 
(\ref{cordoffang}), the corresponding 
perturbative spectral density which enters in Eq.~(\ref{dis}) is:
\beqa
\rho^{(K)}_{\mu}(s,u,Q^2)& = & \frac{3}{2\pi\sqrt\lambda}
 \left\{p_\mu \left[A \left(m_c^2+m_c m_s - 2k\cdot p+p\cdot\pli\right)+
2\pi\left(m_c^2-m_c m_s - k\cdot \pli \right) \right]  \right.\nonumber 
\\
& &+ \pli_{\mu}\left.\left[B \left(m_c^2+m_c m_s - 2k\cdot p+p\cdot\pli
\right)+2\pi\left(k\cdot p-m_c^2\right) \right]\right\} 
\label{sdkoffang}
\eeqa
for $K$ off-shell, where $k\cdot p=\frac{s+m_c^2}{2}$, 
$k\cdot\pli=\frac{u+m_c^2-m_s^2}{2}$, and 
\beqa
\rho^{(D)}_{\mu \nu}(s,u,Q^2)&=&-\frac{3 i}{2\pi\sqrt\lambda} 
\left\{ 
g_{\mu \nu} \left[2\pi\left(m_s^2(m_c-m_s)+m_s\left(k\cdot p+ k\cdot p'-
p\cdot p'\right) -m_ck\cdot p\right)-
2 m_c D\right]\right.  \nonumber \\
&&+p_{\mu} \pli_{\nu} \left[A (m_c+m_s) + B m_s-2 C m_s  -2 \pi m_s 
\right]  \nonumber \\
&&+\pli_{\mu} p_{\nu} \left[A (m_c-m_s) - B m_s-2 C m_s  +2 \pi m_s 
\right]  \nonumber \\
&&-\left. p_{\mu} p_{\nu} 2 m_c F + \pli_{\mu} \pli_{\nu}2  m_c (B-E) 
\right\} 
\label{sddoffang}
\eeqa
for $D$ off-shell, where $k\cdot p=\frac{s+m_s^2-m_c^2}{2}$ and $k\cdot
\pli=\frac{u+m_s^2}{2}$. The definitions of the other quantities are 
the same as for the $D^* D_s K$ vertex, with
\[
\begin{array}{rclrclrcl}
\overline{k}_0&=&\displaystyle{\frac{s+m_c^2}{2\sqrt{s}}},\;\;\;\;&
|\overline{\vec{k}}|&=&\displaystyle{\sqrt{\overline{k}_0^2-m_c^2}},\;\;
\;\;&
\cos\overline{\theta}&=&\displaystyle{\frac{2p'_0\overline{k}_0+m_s^2-
m_c^2-u}{2|\vec{\pli}||\overline{\vec{k}}|}},
\end{array}
\]
for $K$ off-shell, and
\[
\begin{array}{rclrclrcl}
\overline{k}_0&=&\displaystyle{\frac{s+m_s^2-m_c^2}{2\sqrt{s}}},\;\;\;\;&
|\overline{\vec{k}}|&=&\displaystyle{\sqrt{\overline{k}_0^2-m_s^2}},\;\;\;
\;&
\cos\overline{\theta}&=&\displaystyle{\frac{2p'_0\overline{k}_0-m_s^2-u}
{2|\vec{\pli}||\overline{\vec{k}}|}},
\end{array}
\]
for $D$ off-shell.

The phenomenological side of the vertex functions are obtained by 
considering the contributions 
of $D$ and $D_s^*$ mesons to the matrix element in 
Eq.~(\ref{corkoffang}) and $D_s^*$ and $K$ mesons to the matrix 
element in Eq.~(\ref{cordoffang}). We introduce the decay constants 
$f_D$ and $f_{D_s^*}$, which are defined by the following matrix elements:
\beqa
\langle 0|j^{D}|D\rangle&=&\frac{m_D^2}{m_c+m_q}f_D, \label{fd} \\ 
\langle 0|j_{\nu}^{D_s^*}|{D_s^*}\rangle&=&m_{D_s^*}f_{D_s^*} 
\epsilon^*_{\nu},\label{fds*}
\eeqa
where $\epsilon_{\nu}$ is the polarization vector of the $D_s^*$ meson.
The $f_K$ decay constant was already defined in Eq.~(\ref{fk}). Again
 we have to choose one Dirac structure for each case in 
Eqs.~(\ref{sdkoffang})--(\ref{sddoffang}). Following the points discussed 
before, the chosen Dirac structures  are  $\pli_\mu$ for the off-shell $K$, 
and $\pli_\mu\pli_\nu$ for the off-shell $D$. The corresponding 
phenomenological amplitudes in these structures are
\begin{equation}
\Gamma^{(K)ph}(p^2,{\pli}^2,Q^2)= g^{(K)}_{D_s^*DK}(Q^2)
 \frac{f_{D_s^*}f_Df_K m_{D_s^*}m_D^2m_K^2}{m_cm_s
(p^2-m^2_D)({\pli}^2 -m^2_D)(Q^2+m^2_K)}\left(1+\frac{m_D^2+Q^2}
{m_{D_s^*}^2}\right) 
\label{phenkoffang}
\end{equation}
for $K$ off-shell, and 
\begin{equation}
\Gamma^{(D)ph}(p^2,{\pli}^2,Q^2)=g^{(D)}_{D_s^*DK}(Q^2)
\frac{(-2) i f_{D_s^*}f_Df_K m_{D_s^*}m_D^2}
{m_c(p^2-m^2_{D_s^*})({\pli}^2 -m^2_K)(Q^2+m^2_D)}
\label{phendoffang}
\end{equation}
for $D$ off-shell. As in the case of the $D^* D_sK$ vertex, the quark 
condensate does not contribute to the sum rule for these structures.

The procedure to obtain the QCD sum rule is the same used in the case of 
the $D^* D_sK$ vertex studied before. In this case we use the following 
relations between the Borel masses: $M^2/M'^2=m_D^2/m_{D_s^*}^2$ for $K$ 
off-shell and $M^2/M'^2=m_{D_s^*}^2/m_\rho^2$ for $D$ off-shell. The 
values of the parameters used in the calculation of the $D_s^*DK$ vertex 
are given in Table~\ref{param2}, where we have used the relation 
$f_{D_s^*}=f_{D^*}f_{D_s}/f_D$ and the value of $f_{D_s}/f_D$ from Ref.~\cite{fds*} 
in order to obtain the $D_s^*$ decay constant.
\begin{table}[t]
\begin{tabular}{|c|c|c|c|c|c|c|c|c|c|}\hline
$m_q$&$m_s$&$m_c$&$m_K$&$m_{D}$&$m_{D_s^*}$&$f_K$\cite{fk}&$f_{D}$
\cite{fd}&$f_{D_s^*}$\\ 
\hline\hline
0.0&0.13&1.2&0.498&1.87&2.11&0.160&0.200&0.330\\ \hline
\end{tabular}
\caption{Parameters used in the calculation of the QCD sum rule for the 
$D_s^*DK$ vertex. All quantities are in $\GeV$.}\label{param2}
\end{table}

\begin{figure}[t] 
\begin{center}
\epsfig{file=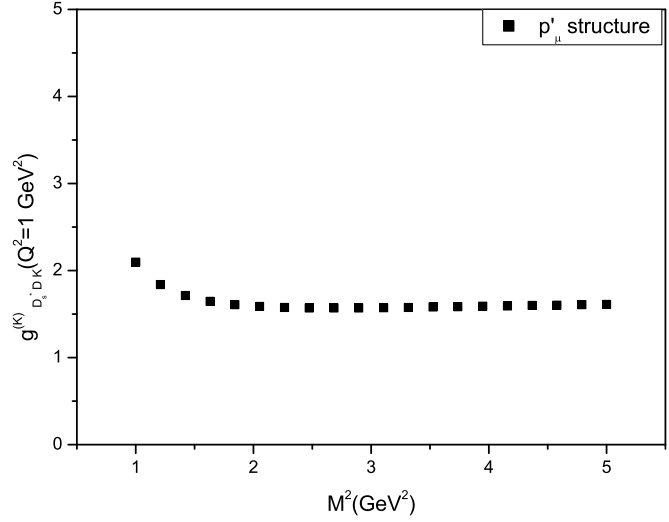}
\caption{Stability of $g^{(K)}_{D^*_s D K}(Q^2=1 \GeV^2)$, as a function of 
the Borel mass $M^2$.}
\label{fig7}
\end{center}
\end{figure}

\begin{figure}[t] 
\begin{center}
\epsfig{file=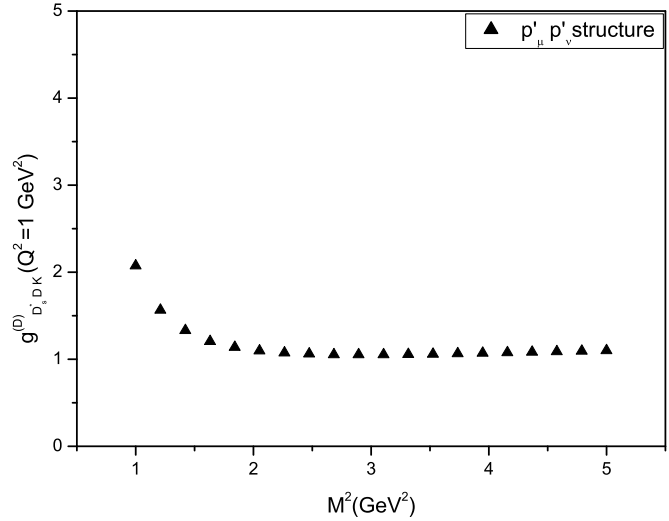}
\caption{Stability of $g^{(D)}_{ D^*_s D K}(Q^2=1 \GeV^2)$, as a
function of the Borel mass $M^2$.}
\label{fig8}
\end{center}
\end{figure}

Using $\Delta_s=\Delta_u = 0.5 \GeV $ for the continuum thresholds 
and fixing $Q^2=1 \GeV^2$, we found a good stability of the 
sum rule for $g_{ D^*_s D K}^{(K)}$, as a function of the Borel mass 
$M^2$, in the interval 
$2< M^2 < 5 \GeV^2$, as can be seen in Fig.~\ref{fig7}. In the case of 
$g_{D^*_s D K}^{(D)}$, the interval for stability is also
$2 <M^2 < 5 \GeV^2 $, as can be seen in Fig.~\ref{fig8}. 

\begin{figure}[t] 
\begin{center}
\epsfig{file=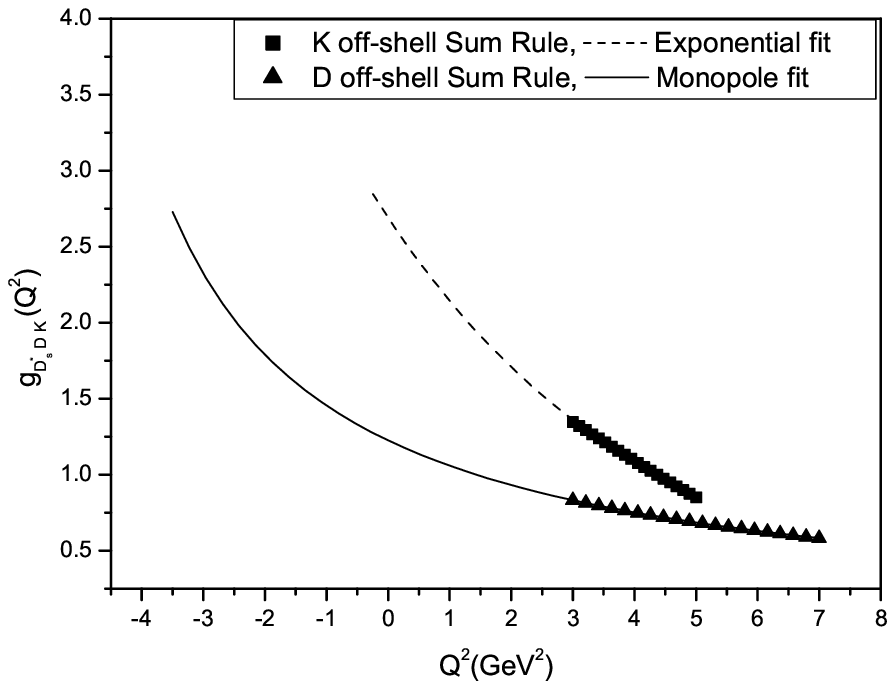}
\caption{$g^{(K)}_{D^*_s D K}$ (squares) and 
$g^{(D)}_{D^*_s D K}$ (triangles) form factors as a function of
$Q^2$ from the QCDSR calculation of this work. The dashed (solid) line 
corresponds to the exponential (monopole) parametrization of the QCDSR 
results for each case.}
\label{fig9}
\end{center}
\end{figure}

Fixing $\Delta_s=\Delta_u=0.5 \GeV$ and $M^2=3 \GeV ^2$ in both cases,
we calculate the momentum dependence of the form factors which are shown 
in  Fig.~\ref{fig9}. The squares corresponds to the 
$g_{D^*_s D K}^{(K)}(Q^2)$ form factor in the  interval where the 
sum rule is valid. The triangles are the result of the sum rule for the 
$g_{D^*_s D K}^{(D)}(Q^2)$ form factor. In the case when the $K$ 
meson is off-shell, our numerical results can be 
parametrized by an exponencial function (dashed curve in Fig.~\ref{fig9}):
\begin{equation}
g_{D^*_s D K}^{(K)}(Q^2)= 2.69 \; e^{-\frac{Q^2}{4.39}}.
\label{expang}
\end{equation}
The coupling constant was obtained as the value of the form factor at  
$Q^2= -m^2_K$. In this case the resulting coupling constant is
\begin{equation}
g_{D^*_s D K}^{(K)}= 2.87.  \label{gexpang}
\end{equation}
In the case when the $D$ meson is off-shell, the sum rule result is
represented by the triangles in 
Fig.~\ref{fig9}, and they can be parametrized by a monopole formula
(solid line in the figure):  
\begin{equation}
g_{D^*_s D K}^{(D)}(Q^2)=\frac{7.78}{Q^2+6.34},
\label{monoang}
\end{equation}
giving the following coupling constant, obtained at the $D$ pole:
\begin{equation}
g_{D^*_s D K}^{(D)}= 2.72.\label{gmonopolenos}
\end{equation}

\begin{figure}[t] 
\begin{center}
\epsfig{file=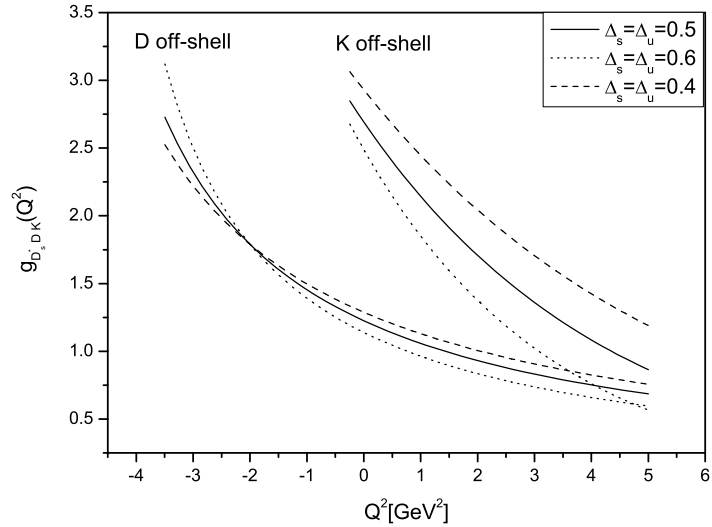}
\caption{Dependence of the form factor with the continuum threshold, for 
$K$ and $D$ off-shell cases. The dotted line
correponds to $\Delta_s=\Delta_u=0.4~\GeV$, the solid corresponds to 
$\Delta_s=\Delta_u=0.5~\GeV$ and the dashed corresponds
to $\Delta_s=\Delta_u=0.6~\GeV$.}
\label{fig10}
\end{center}
\end{figure}

Studing the dependence of our results with the continuun threshold, for 
$\Delta_{s,u}$ varying in the interval  $0.4 \le \Delta_{s,u}\le 0.6~\GeV$, as can be seen in
 Fig.~{\ref{fig10}}, we obtain the following values, with errors, for the couplings in each case: 
 $g_{D^*_s D K}^{(K)}= 2.87 \pm 0.19 $ and $g_{D^*_s D K}^{(D)}= 2.72 \pm 0.31$.

Concluding, we have estudied the form factors and coupling constants of 
$D^* D_s K$ and 
$D_s ^* D K$ vertices in a QCD sum rule calculation. For each case we 
have considered two particles off-shell, the lightest and one of the 
heavy ones: the $K$ and $D_s$ mesons for the $D^*D_sK$ vertex, and the 
$K$ and $D$ mesons for the $D_s ^* D K$ vertex. 
In the two situations, the off-shell particles give compatible results 
for the coupling constant in each vertex.
 The results are:
\begin{eqnarray}
g_{D^* D_s K}&=&3.02 \pm 0.14 \label{g1}\\
g_{D_s ^* D K}&=& 2.84 \pm 0.31 \label{g2}
\end{eqnarray} 

We can compare our result with the prediction of the exact SU(4) 
symmetry \cite{lin,oh,regina}, which would 
give the following relation among these numbers \cite{regina}: $g_{D^* D_s K}=
g_{D_s ^* D K} = 5.$
Eqs.~(\ref{g1}) and (\ref{g2}) shows that the coupling constants 
in the vertices $D^* D_s K$ and $D_s ^* D K$ are consistent one with the  
other, but that they are relatively 
far from the value given by the SU(4) symmetry in the cited reference. Therefore, 
we conclude that the  SU(4) symmetry is broken by 
approximately 40\% in the calculation performed here. 
We can also extract the cutoff parameter, 
$\Lambda$, from the paramentrizations in 
Eqs.~(\ref{expnos}) and (\ref{expang}) for $K$ off-shell, Eq.~(\ref{mononos}) 
for $D_s$ off-shell and Eq.~(\ref{monoang}) for $D$
off-shell. We get $\Lambda \approx 2.07~\GeV$ for the $K$ meson off-shell , 
$\Lambda \approx 2.61~\GeV$
for the $D_s$ meson off-shell, and $\Lambda \approx 2.51~\GeV$
for the $D$ meson off-shell. Comparing the values of the cutoffs, we see 
that the form factor is harder if the off-shell meson is heavier, implying 
that the size of the vertex depends on the mass of the exchanged meson: 
the heavier is the meson, the more as a point like particle is its 
behavior when probing the target, as observed in 
Refs.~\cite{ddpir,ddrho,psid*d*,matheus1}.

\vspace{0.5cm} 

\underline{Acknowledgements}: 
This work has been supported by CNPq and FAPESP.

\end{document}